\documentclass[]{aa} 

\usepackage{graphicx}
\usepackage{txfonts}
\usepackage{xcolor}

\usepackage[breaklinks, colorlinks, citecolor=blue, linkcolor=blue]{hyperref}

\DeclareMathOperator*{\argmin}{arg\,min}
\newcommand{\norm}[1]{\left\lVert#1\right\rVert}
\newcommand{\vect}[1]{\boldsymbol{\mathbf{#1}}}
\begin{document} 

   \title{Planet cartography with neural learned regularization}

   \author{A. Asensio Ramos \and E. Pall\'e}

   \institute{Instituto de Astrof\'isica de Canarias (IAC), Avda V\'ia L\'actea S/N,
		38200 La Laguna, Tenerife, Spain\\
              \email{andres.asensio@iac.es}
         \and
             Departamento de Astrof\'isica, Universidad de La Laguna, 38205 La Laguna, Tenerife, Spain\\
              }

   \date{Received ; accepted }

 
  \abstract
   {}
   {Finding potential life harboring exo-Earths is one of the aims of exoplanetary
    science with the use of future telescopes. Detecting signatures of life in exoplanets will 
    likely first be accomplished by determining the bulk composition of the planetary atmosphere 
    via reflected/transmitted spectroscopy. However, a complete understanding of the habitability conditions 
    will surely require mapping the presence of liquid water, continents and/or clouds.
    Spin-orbit tomography is a technique that allows us to obtain maps of the surface of
    exoplanets around other stars using the light scattered by the planetary surface.}
   {We leverage the enormous potential of deep learning and propose
    a mapping technique for exo-Earths in which the regularization is learned from
    mock surfaces. The solution of the inverse mapping problem is posed as a deep
    neural network that can be trained end-to-end with suitable training data. Since we
    still lack observational data of the surface albedo of exoplanets, we propose in this
    work to use methods based on the procedural generation of planets, inspired
    by what we found on Earth. We also consider mapping the recovery of surfaces and the presence
    of persistent cloud in cloudy planets, a much more challenging problem.}
   {We show that a reliable mapping can be carried out with our approach, producing
   very compact continents, even when using single passband observations. More importantly, if 
   exoplanets are partially cloudy like the Earth is, we show that one can potentially map the distribution of
   persistent clouds that always occur on the same position on the surface (associated 
   to orography and sea surface temperatures) together
   with non-persistent clouds that move across the surface. This will become the first test one can 
   perform on an exoplanet for the detection of an active climate system. For small rocky planets 
   in the habitable zone of their stars, this weather system will be driven by water, and the 
   detection can be considered as a strong proxy for
   truly habitable conditions. 
}
   {}


   \maketitle
%
\section{Introduction}

One of the most promising goals in the field of exoplanets is to reach the technical capability to characterize 
the bulk composition of rocky exoplanet atmospheres. In particular, for those small planets within the 
habitable zone of their host star, which will in turn open the search for the combination of gaseous species 
that could be classified as a biomarker \citep{DesMarais2002}. With more than 4000 planets detected to date, the number of small 
rocky planets is also large, and it continues to increase thanks to space missions like Kepler and 
TESS as 
well as ground-based photometric and radial velocity surveys. The majority of this small rocky planets, 
however, transit around M-type stellar hosts \citep{Hardegree2019}. This is good news as these small stars offer a greater chance 
for planetary characterization owing to a reduced star/planet contrast ratio. In the coming future, ultra-stable 
high resolution spectrographs, such as ESPRESSO, mounted on large aperture telescopes (ELTs) will allow the detection of 
small rocky planets around K and G type stars.

The initial characterization of small rocky planets will probably be allowed by transmission spectroscopy of 
targets around M-type stars \citep{Palle2011, Snellen2013}, although whether those planets can sustain an atmosphere is still open to 
debate, and more so the possibility of these planets to host life \citep{Tarter2007, Luger2015}. On the longer term, 
direct imaging of habitable zone exoplanets will be accomplished by space-base coronographic or interferometric techniques \citep{Leger1996}. This expansion into reflected light studies is crucial 
for our understanding of planetary atmospheres (and possible biomarkers) in the solar neighborhood, as for 
an Earth-Sun twin the transit probability is only 0.5\%, meaning that only one of the closest 200 earths-like 
planets will transit its host star.

Characterizing planets via reflected light posses a technical challenge, but it also offers the 
possibility of a more in-depth characterization. For example, we will be sensitive to light reflected 
from the surface, and measure the bulk composition of the full atmosphere, as opposed to transmission 
spectroscopy where only the higher levels of the atmosphere are probed \citep{Palle2009}. The initial characterization 
of a planetary atmospheres with reflected light will surely come from single-epoch spectroscopy, which 
will allow the detection of the bulk composition of the atmosphere, searching for species such 
as H$_2$O, CO$_2$, or O$_2$, which might be enough to classify the planet as inhabitable/inhabited \citep{Meadows2018}. 
However, the temporal variability information introduced by cloud-tops and surface reflectance is 
very rich and can provide crucial additional details on the existence of continental land masses 
from single or multi-color photometry \citep{Ford2001}. When clouds come into the picture, retrieving an 
surface map is not so straightforward, but with enough data it can provide insights into the 
existence of a dynamic weather system \citep{2008ApJ...676.1319P}. 

Mapping the surface of exoplanets around other stars using the light 
scattered by the planetary surface was proposed by \cite{2010ApJ...720.1333K}.
Under some approximations, the light scattered by the planet can be linearly
related to the albedo properties of the surface. Mapping the surface albedo
by solving the inverse problem turns out to be, like many other linear
and nonlinear problems, ill-posed. There is not enough information on the light
curve to fully constrain the solution and some kind of regularization has to be used. 
\cite{2011ApJ...739L..62K} formulated the inverse problem, named spin-orbit
tomography (SOT) and proposed a solution based on the Tikhonov regularization. 
The general idea is that if one observes a spinning planet orbiting around a star 
for a sufficiently long time and with enough time cadence, one can reconstruct
the surface albedo from the time modulation of the light curve. This was indeed
nicely demonstrated by \cite{2012ApJ...755..101F}, who showed that approximate albedo maps can
be obtained for planets with different orbital configurations.

After the initial idea of SOT, other approaches to the solution of the inverse
problem have been suggested. \cite{2018AJ....156..146F} proposed to use a Gaussian Process (GP)
as a prior for the surface albedo map in the pixel domain and deal with the inversion
in the Bayesian framework. This allowed \cite{2018AJ....156..146F} to produce 2D mappings
with uncertainties and show that a planet can be mapped with a very limited
amount of observations. This work has been recently extended by \cite{2020ApJ...900...48K}
to also propose a GP in time as a prior of the time evolution of
the surface. This can be used for planets with moving surfaces or with
clouds that evolve in time. \cite{2019AJ....157...64L} developed \texttt{starry}, a 
computer code which uses spherical harmonics as a basis set of the albedo surface of a planet. 
Apart from a very compact representation of the surface albedo, this representation 
also leads to an efficient acceleration of the computations. This allowed \cite{2019arXiv190312182L} to
map a mock Earth using synthetic TESS data.
\cite{2019AJ....158..246B} later proposed an Occamian approach for the solution
of the inverse problem, which is essentially equivalent to solving the linear
inverse problem using the singular value decomposition (SVD). Recently,
\cite{2020ApJ...896...22A} leveraged the recent theories of compressed sensing 
and sparse reconstruction \citep{candes06} to show that such sparse reconstructions (also
discussed in the next section) lead to an improved mapping of the surface,
with better constrained continents. More recently, \cite{2020ApJ...894...58K} used non-negative matrix 
factorization to show how to map exoplanets using different colors. 

All previous approaches are based on some form of mathematical regularization. While Tikhonov
regularization tends to generate smooth maps, sparse regularization produces a better
mapping with more defined continents. GP priors (both spatial and temporal) open up the
possibility of using Bayesian inference but sometimes produce overly smooth results.
In this paper we leverage the enormous potential of deep learning \citep{Goodfellow-et-al-2016} 
and propose a mapping technique for exo-Earths in which the regularization is learned
from simulations. The solution of the inverse mapping problem is posed as a deep neural network that
can be trained end-to-end with suitable training data. Since we still
lack observational data of the surface albedo of exoplanets, our results depend on our ability
to generate mock surfaces. We propose in this work to use methods based on the procedural 
generation of planets that are inspired by the appearance of the Earth's real surface and clouds properties.

\begin{figure}
    \centering
    \includegraphics[width=0.5\columnwidth]{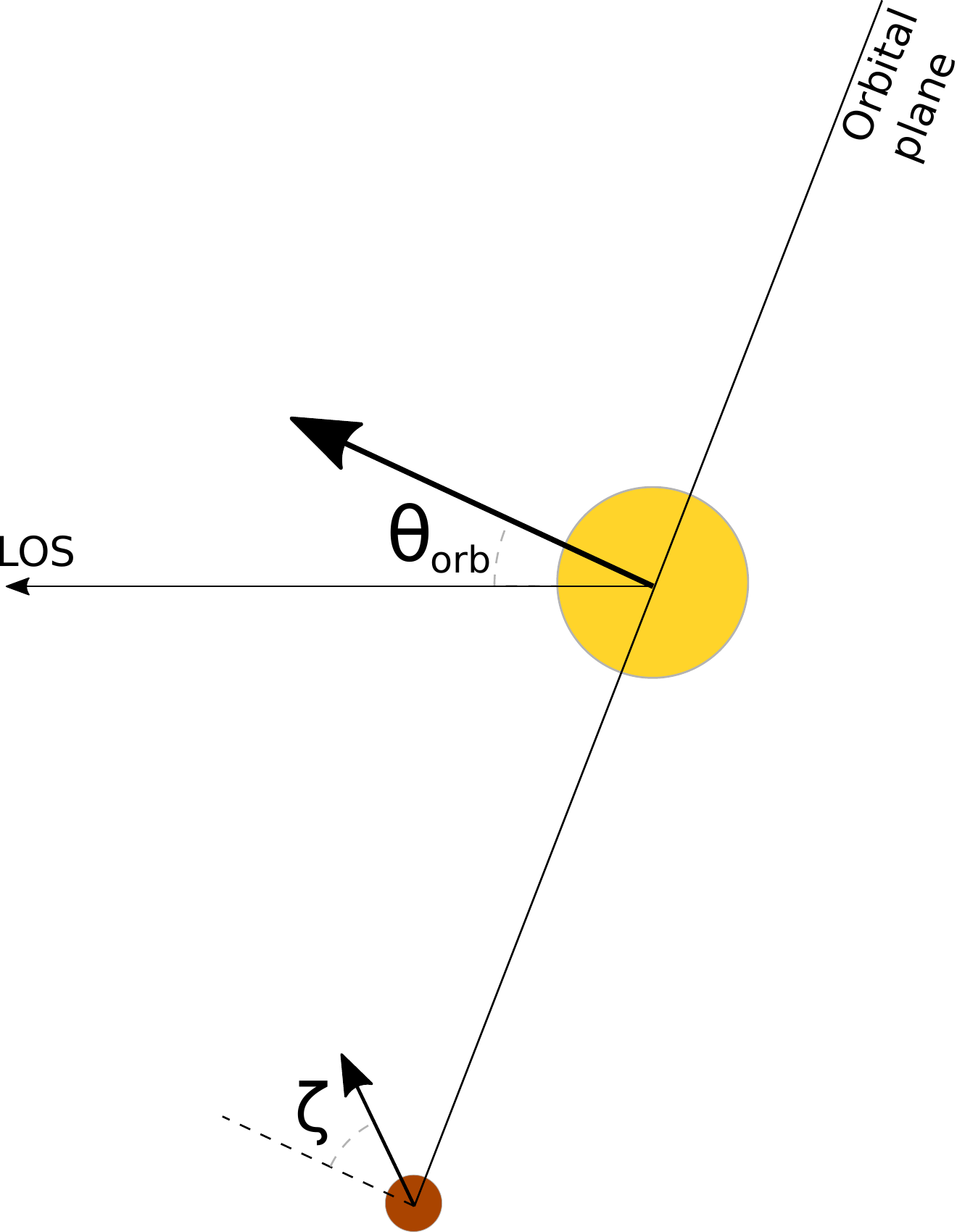}
    \caption{Schematic illustration of the orbital configuration. The orbital plane
    is defined by the angle $\theta_\mathrm{orb}$, while the planet spins forming an obliquity angle
    $\zeta$ with respect to the orbital plane normal.}
    \label{fig:geometry}
\end{figure}

\section{Mapping cloudless planets}
\subsection{Formulation of the problem}
The ratio between the light scattered by the surface of an exoplanet orbiting around a certain
star and the incoming flux of the star is, in general, a complex function of the
geometry of scattering and the properties of the scatterers. In the
single scattering approximation and assuming scattering is isotropic (Lambertian), this
function only depends on geometry.

For a given time $t_i$, the integrated stellar light reflected from an exoplanet is given by the following
linear relation when the planetary surface is conveniently discretized:
\begin{equation}
    d_i = \sum_j \Phi_{ij} m_j,
    \label{eq:model}
\end{equation}
where $d_i$ is the relative flux observed at time $t_i$ and $m_j$ is the $j$-th surface element albedo of the 
discretized planetary surface. The matrix elements $\Phi_{ij}$ are purely geometrical, and depend
on the relative position of the $j$-th surface element of the planet, the  
star and the observer at time $t_i$. In matrix form, and assuming a realistic 
observation, we have:
\begin{equation}
    \mathbf{d} = \boldsymbol{\Phi} \mathbf{m} + \vect{\epsilon},\label{eq:generative}
\end{equation}
where $\vect{\epsilon}$ is Gaussian noise with zero mean and variance $\sigma_n$. The matrix $\boldsymbol{\Phi}$,
also known as the geometric kernel, 
depends on the following set of parameters of the planet (see Fig. \ref{fig:geometry}): the rotation period ($P_\mathrm{rot}$), 
the orbital period ($P_\mathrm{orb}$), the initial rotational and orbital phases ($\phi_\mathrm{rot}$ and
$\phi_\mathrm{orb}$, respectively), the orbit inclination $\theta_\mathrm{orb}$ and the obliquity of
the planet $\zeta$.

When observations are available, it is possible to invert the previous linear
relation. Given that the problem is ill-defined, the solution is often regularized, so that
one needs to solve the following maximum a-posteriori problem:
\begin{equation}
\argmin_m \frac{1}{2} \norm{\mathbf{d} - \boldsymbol{\Phi} \mathbf{m}}^2_2 + \lambda R(\mathbf{m})
\label{eq:problem},
\end{equation}
where the log-likelihood (first term) is a direct consequence of the assumption of uncorrelated Gaussian noise,
while $R(\mathbf{m})$
plays the role of a convex log-prior that is used to regularize the result. Several different 
options have been used in the literature, as described in the Introduction. The first one is Tikhonov regularization, 
for which $R(\mathbf{m})=\norm{\mathbf{m}-\hat{\mathbf{m}}}^2_2$. This regularization pushes the solution
to $\hat{\mathbf{m}}$ when no information is available. More recently, sparse 
regularization (or $\ell_1$ regularization) has also been considered directly on the albedo surface, so that
$R(\mathbf{m})=\norm{\mathbf{m}}_1$\footnote{We note that the $\ell_1$ norm of
a vector is given by $\norm{\mathbf{x}}_1 = \sum_i |x_i|$.}. 
It is also interesting to point out that this regularization 
can also be applied in a transformed domain with $R(\mathbf{m})=\norm{\mathbf{W} \mathbf{m}}_1$, 
where $\mathbf{W}$ is often a linear orthonormal transformation with a fast transform operator, like
Fourier or wavelet. \cite{2020ApJ...896...22A} assumed sparsity on the pixel space (large regions
of the planet have small albedos compatible with zero), but it might well be that 
the solution is ever sparser in a transformed domain.

\begin{figure*}
    \centering
    \includegraphics[width=\textwidth]{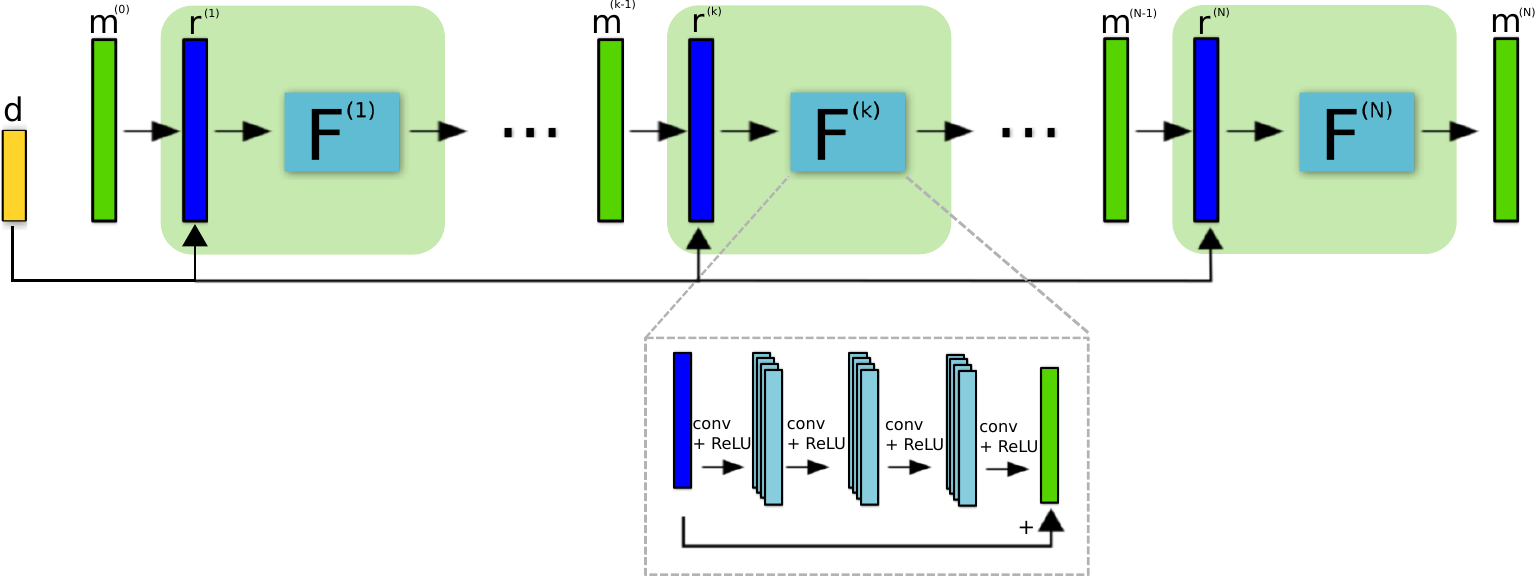}
    \caption{Architecture of the deep neural network that carries out the unrolled iterative
    reconstruction for $N$ steps. Each green phase plate defines the application of Eqs. (\ref{eq:step1}) and
    (\ref{eq:step2}). The internal structure of the $F$ operator is shown
    in the inset. These operators are made of a train of convolutional layers with ReLU activation functions
    that generate $N_f$ channels. The last layer , followed by a 
    ReLU activation function and another linear layer.}
    \label{fig:architecture}
\end{figure*}

\subsection{Deep neural network regularized reconstruction with loop unrolling}
A conceptually simple, often fast, and successful method for the solution of convex regularized problems is
the iterative shrinkage thresholding algorithm \citep[ISTA;][]{ista09}. ISTA is a first order proximal
algorithm \citep{parikh2013proximal} for the optimization of the following type of problems:
\begin{equation}
    \argmin_\mathbf{x} f(\mathbf{x}) + g(\mathbf{x}),
\end{equation}
where $f(\mathbf{x})$ is differentiable and $g(\mathbf{x})$ is a convex function. The solution is obtained by the
iterative application of the following steps:
\begin{align}
    \mathbf{r}^{(k)} &= \mathbf{x}^{(k-1)} - \rho \nabla f(\mathbf{x})\label{eq:ista}\\
    \mathbf{x}^{(k)} &= \texttt{prox}_{g}(\mathbf{r}^{(k)}) = \argmin_{\mathbf{y}} \frac{1}{2} \norm{\mathbf{y} - \mathbf{r}^{(k)}}^2_2 + g(\mathbf{y})\label{eq:proximal_ista_general},
\end{align}
where $\rho$ is a step size that needs to be properly tuned. The ISTA algorithm (and fundamentally 
all algorithms for the solution of regularized problems) crucially depend on the availability of a fast way 
to solve subproblem (\ref{eq:proximal_ista_general}), which is
known as the proximal operator. Fortunately, it is possible to find a closed form for
the proximal operator for certain $g(\mathbf{x})$ functions. 
Of special interest are the case of $\ell_1$ and Tikhonov regularizations. The 
proximal operator for the $\ell_1$ regularization, where $g(\mathbf{x})=\norm{\mathbf{W} \mathbf{x}}_1$, is given by:
\begin{equation}
    \texttt{prox}_{\lambda \norm{\mathbf{W} \cdot}_1}(\mathrm{x}) = \mathbf{W}^T \mathrm{soft}(\mathbf{W} \mathbf{x},\lambda), \label{eq:proximal_l1}
\end{equation}
where the soft thresholding (or shrinkage) operator is given by:
\begin{equation}
    \mathrm{soft}(x,\lambda) = \mathrm{max} \left( x - \lambda,0 \right) \frac{x}{\left| x \right|}.
\end{equation}
The proximal operator for the Tikhonov regularization, in which $g(\mathbf{x})=\norm{\mathbf{x}}_2$, is given by
\begin{equation}
    \texttt{prox}_{\lambda \norm{\cdot}_2}(\mathbf{x}) = \mathrm{max} \left( 1- \frac{\lambda}{\norm{\mathbf{x}}_2},0 \right) \mathbf{x}. \label{eq:proximal_tikhonov}
\end{equation}

Particularizing the ISTA method to Eq. (\ref{eq:problem}), the iterative 
scheme is given by:
\begin{align}
    \mathbf{r}^{(k)} &= \mathbf{m}^{(k-1)} - \rho \boldsymbol{\Phi}^T \left( \boldsymbol{\Phi} \mathbf{m}^{(k-1)} - \mathbf{d} \right) \label{eq:gradient}\\
    \mathbf{m}^{(k)} &= \texttt{prox}_{\lambda R}(\mathbf{r}^{(k)}) = \argmin_{x} \frac{1}{2} \norm{\mathbf{x} - \mathbf{r}^{(k)}}^2_2 + \lambda R(\mathbf{x})\label{eq:proximal},
\end{align}
where the step size $\rho$ has an optimal value given by $\rho \leq (2 \tau)^{-1}$, where $\tau$ is
the largest eigenvalue of the $\boldsymbol{\Phi}^T \boldsymbol{\Phi}$ matrix.

We propose to solve the regularized problem by leveraging 
deep neural networks to impose the regularization. Instead of 
explicitly defining the regularization function $R(\mathbf{m})$, we directly
model its effect as a proximal projection with the aid of a neural network, so we substitute 
Eq. (\ref{eq:proximal}) by:
\begin{equation}
\mathbf{m}^{(k)} = \texttt{prox}_{\lambda R}(\mathbf{r}^{(k)})  = F^{(k)}(\mathbf{r}^{(k)})
 \label{eq:proximal_F}.
\end{equation}
where the $F^{(k)}(\mathbf{x})$ are mappings defined by convolutional neural networks whose architecture we discuss below. 
The $F^{(k)}(\mathbf{x})$ operators can then be seen as \emph{learned denoisers} that generalize the concept of
a proximal projection operator \citep[e.g.,][]{venkatakrishnan13,zhang18,wang19}. Such iterative methods
are commonly known as \emph{plug-and-play methods}.


We follow \cite{gregor-icml-10} and unroll $N$ iterations of Eqs. (\ref{eq:gradient}) and (\ref{eq:proximal_F})
as a deep neural network and train the network end-to-end. The loop unrolling also allows us
to learn optimal values for $\rho$ used in the ISTA iteration. We also allow
them to be different at all iterations, not like ISTA in which they are kept constant. For clarity, we repeat here
the two steps that need to be iterated for $k=1,\ldots,N$, starting from an initial solution $\mathbf{m}^{(0)}=0$:
\begin{align}
    \mathbf{r}^{(k)} &= \mathbf{m}^{(k-1)} - \rho \rho^{(k)} \boldsymbol{\Phi}^T \left( \boldsymbol{\Phi} \mathbf{m}^{(k-1)} - \mathbf{d} \right) \label{eq:step1}\\
    \mathbf{m}^{(k)} &= F^{(k)} (\mathbf{r}^{(k)}) \label{eq:step2}.
\end{align}
The learned quantities are forced to fulfill $\rho^{(k)} \sim 1$ and are used to allow our trained unrolled
scheme to use step sizes different than the optimal one given by $\rho$.
We have found that the chosen initial solution gives stable reconstructions, although one
can also use other potentially interesting options like $\mathbf{m}^{(0)}=\boldsymbol{\Phi}^T \mathbf{d}$.

The general structure of the unrolled deep network is displayed in Fig. \ref{fig:architecture}. The inputs 
to the network are the vector of observations $\mathbf{d}$ and the initial solution $\mathbf{m}^{(0)}$. Thess 
vectors are then used to compute $\mathbf{r}^{(k)}$, which is then transformed by 
the non-linear neural network to finally produce a 
new estimated solution. This process is iterated $N$ times. The detail of the neural networks is 
shown as an inset in Fig. \ref{fig:architecture}. The specifics of these networks 
is discussed in the following.

\subsection{Denoisers}
Several options are available for the layers inside the $F$ operators.
The first obvious option is to use a fully connected layer to map the HEALPix pixelization, seen
as a one-dimensional vector, to intermediate representations. We quickly discarded this option 
because the number of free parameters of the network 
increases very fast with the number of pixels in the sphere. The second option is to use convolutional layers, which are very
efficient in the number of parameters. Our first option is what we term an \emph{1D regularizer}, that
uses convolutional layers acting on the HEALPix pixelization seen as a one-dimensional vector.
This option does not exploit any specific spatial correlation in the surface of the
exoplanet. As indicated in Fig. \ref{fig:architecture}, we utilize four 1D convolutional 
layers with kernels of size 3 producing the sequence
$(1,N_s) \to (32,N_s) \to (32,N_s) \to (32,N_s) \to (1,N_s)$, where the first index indicates the number
of channels and the second index indicates the number of pixels on the surface. 
We unroll $N=15$ iterations, with a total number of $\sim$95k free parameters.
An arguably more appropriate option, that we term as an \emph{2D regularizer}, is to use 
convolutional layers defined on the
surface of the sphere using the HEALPix pixelization. We take advantage of the idea developed by
\cite{2019A&A...628A.129K}. This option has the interesting
property of exploiting the spatial information on the sphere to produce an intermediate
representation. In this case we also use 32 kernels of size $3 \times 3$ in the
intermediate layers, resulting in a total of 285k free parameters. We caution that
the solution of \cite{2019A&A...628A.129K} for implementing convolutions on the sphere is arguably not
the most optimal, producing artifacts that we discuss in the following sections. It is worth exploring other 
options for convolutions in the sphere \citep[e.g.,][]{2019A&C....27..130P}.

\subsection{Tikhonov regularization}
For comparison with our results we consider the Tikhonov-regularized case. We solve it using
ISTA with the corresponding proximal projection operator, given by Eq. (\ref{eq:proximal_tikhonov}):
\begin{align}
    \mathbf{r}^{(k)} &= \mathbf{m}^{(k-1)} - \rho \boldsymbol{\Phi}^T \left( \boldsymbol{\Phi} \mathbf{m}^{(k-1)} - \mathbf{d} \right) \label{eq:gradient_tikhonov}\\
    \mathbf{m}^{(k)} &=  \mathrm{max} \left( 1- \frac{\lambda}{\norm{\mathbf{r}^{(k)}}_2},0 \right) \mathbf{r}^{(k)}, 
    \label{eq:tikhonov}
\end{align}
starting from $\mathbf{m}^{(0)}=0$.

\begin{figure*}
    \centering
    \includegraphics[width=0.9\textwidth]{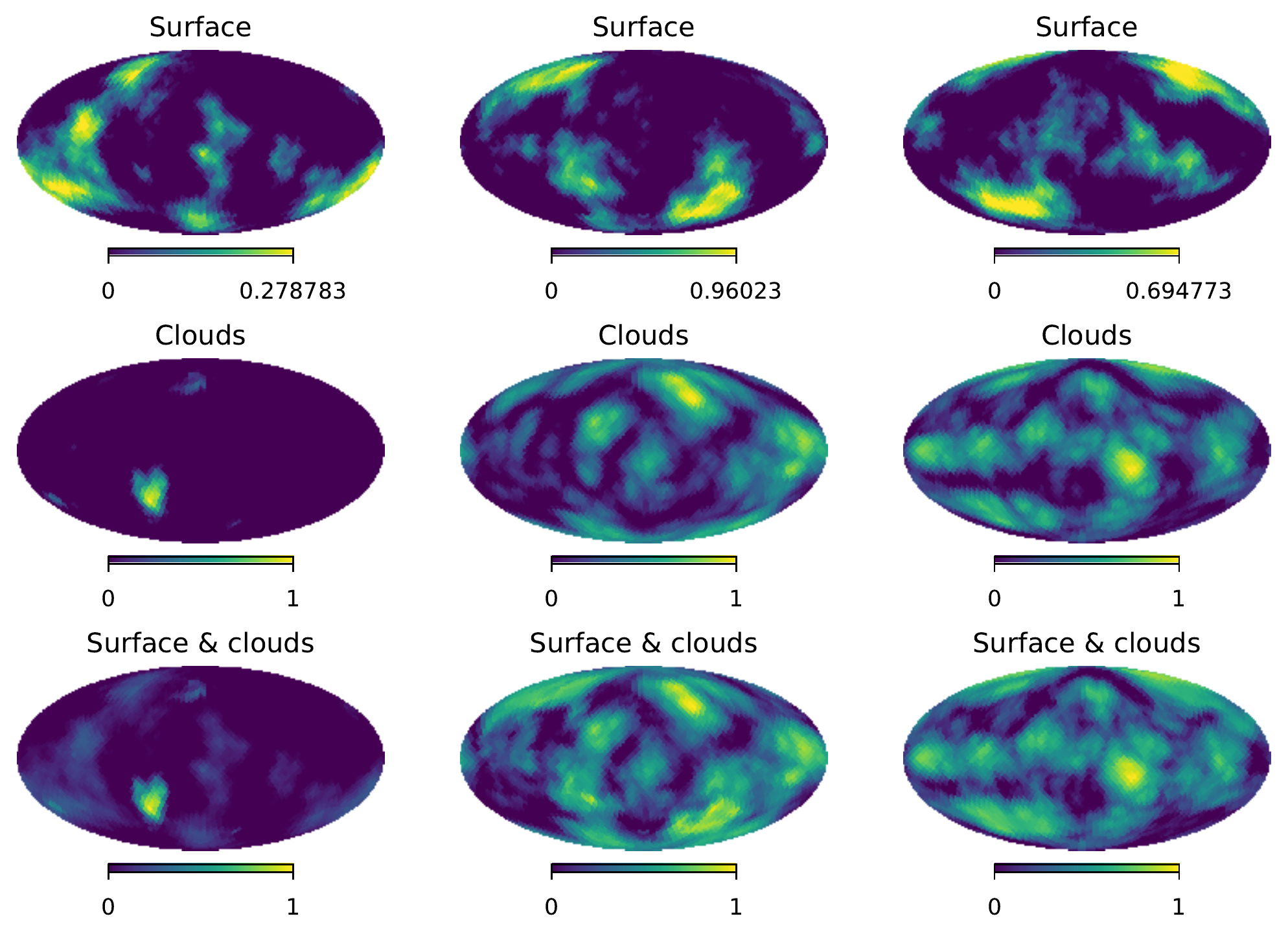}
    \caption{Samples of mock planetary (upper row) and cloud (middle row) albedo surfaces that are part of the training set.
    The combination of the two is shown in the lower row.}
    \label{fig:planets}
\end{figure*}

\section{Mapping cloudy planets}
Mapping cloudy planets is a much more challenging problem than mapping cloudless planets. 
If exoplanets are similar to the Earth, 
one can potentially find persistent clouds that always occur on the same position on the surface together
with non-persistent clouds that form or move across the surface \citep[e.g.,][]{2008ApJ...676.1319P}. Persistent 
clouds can hardly be distinguished from the surface albedo, but non-persistent clouds make the recovery of the surface
much more difficult. Our aim here is to show that this mapping can be carried out
even when using single passband observations, although the mapping can potentially
improve if multi-band observations are considered.

We consider two models for the resulting albedo in a planet with clouds. For the first one, let us
assume, for a discretized element in the surface of the planet, that $m_s$ 
is the surface albedo while $m_c$ is the cloud albedo. We make the assumption that
the clouds are kept stationary during a single epoch of observations to reduce the
dimensionality of the solution space. We also make the assumption that clouds 
have a fairly low absorption in the visible while scattering is dominant. If the stellar illumination arriving at the 
atmosphere of the planet is $F_0$, the amount of light reflected by the cloud would be $F_0 m_c$. 
A fraction $F_0 (1-m_c)$ would pass downward and reflect off the planetary surface. The amount 
of reflected light would then be $F_0 (1-m_c) m_s$. On its return upward, a fraction $m_c$
is again dispersed and a fraction $(1-m_c)$ would pass to the external layers of the atmosphere. 
Therefore, from this amount of light that comes out again once can infer the final albedo of
the planet, which is $m=m_c + m_s (1-m_c)^2$. By plugging this expression into Eq. (\ref{eq:model}) for
all surface elements, one ends up with the following generative model for the observed
flux at epoch $i$ from a total of $N_e$ epochs:
\begin{equation}
    \mathbf{d}_i = \boldsymbol{\Phi}_i \left( \mathbf{m}_{c;i} + \mathbf{m}_s \odot (1-\mathbf{m}_{c;i})^2 \right)  + \vect{\epsilon},\label{eq:generative_clouds}
\end{equation}
where $\odot$ is the Hadamard (or elementwise) product. We note that the problem is still linear
in the albedo of the planetary surface if the cloud albedo is known. On the other contrary, the
problem is already non-linear in the cloud albedo. We note that $\mathbf{m}_s$ 
is common to all epochs and plays the role of $\mathbf{m}$ in the cloudless planet.
A direct application of Eqs. (\ref{eq:ista}) and (\ref{eq:proximal_ista_general}) results in the following iterative scheme:
\begin{align}
    \mathbf{r}_s^{(k)} &= \mathbf{m}_s^{(k-1)} - 2 \rho \rho^{(k)} \sum_{i=1}^{N_e} \mathbf{t}_{2;i} \odot \mathbf{t}_{1;i} \label{eq:step1}\\
    \mathbf{m}_{c;i}^{(k)} &= \mathbf{m}_{c;i}^{(k-1)} - 2 \rho \rho^{(k)} \left( \mathbf{t}_{2;i} - 2 \mathbf{t}_{2;i}
    \odot \mathbf{m}_s^{(k-1)} \odot \mathbf{t}_{0;i} \right)\\
    \mathbf{m}_s^{(k)} &= F^{(k)} (\mathbf{r}_s^{(k)}) \label{eq:step2}.
\end{align}
where
\begin{align}
    \mathbf{t}_{0;i} &= 1-\mathbf{m}_{c;i}^{(k-1)}\\
    \mathbf{t}_{1;i} &= \mathbf{t}_{0;i} \odot \mathbf{t}_{0;i}\\
    \mathbf{t}_{2;i} &= \boldsymbol{\Phi}_i^T \left[ \boldsymbol{\Phi}_i \left(\mathbf{m}_{c;i}^{(k-1)} + 
    \mathbf{t}_{1;i} \odot \mathbf{m}_{s}^{(k-1)} \right) - \mathbf{d} \right].
\end{align}
The $F^{(k)}$ operators are trained by unrolling the previous iterative scheme $N$ times
and dealing with them as an end-to-end neural network, as before.

\begin{figure*}
    \centering
    \includegraphics[width=0.9\textwidth]{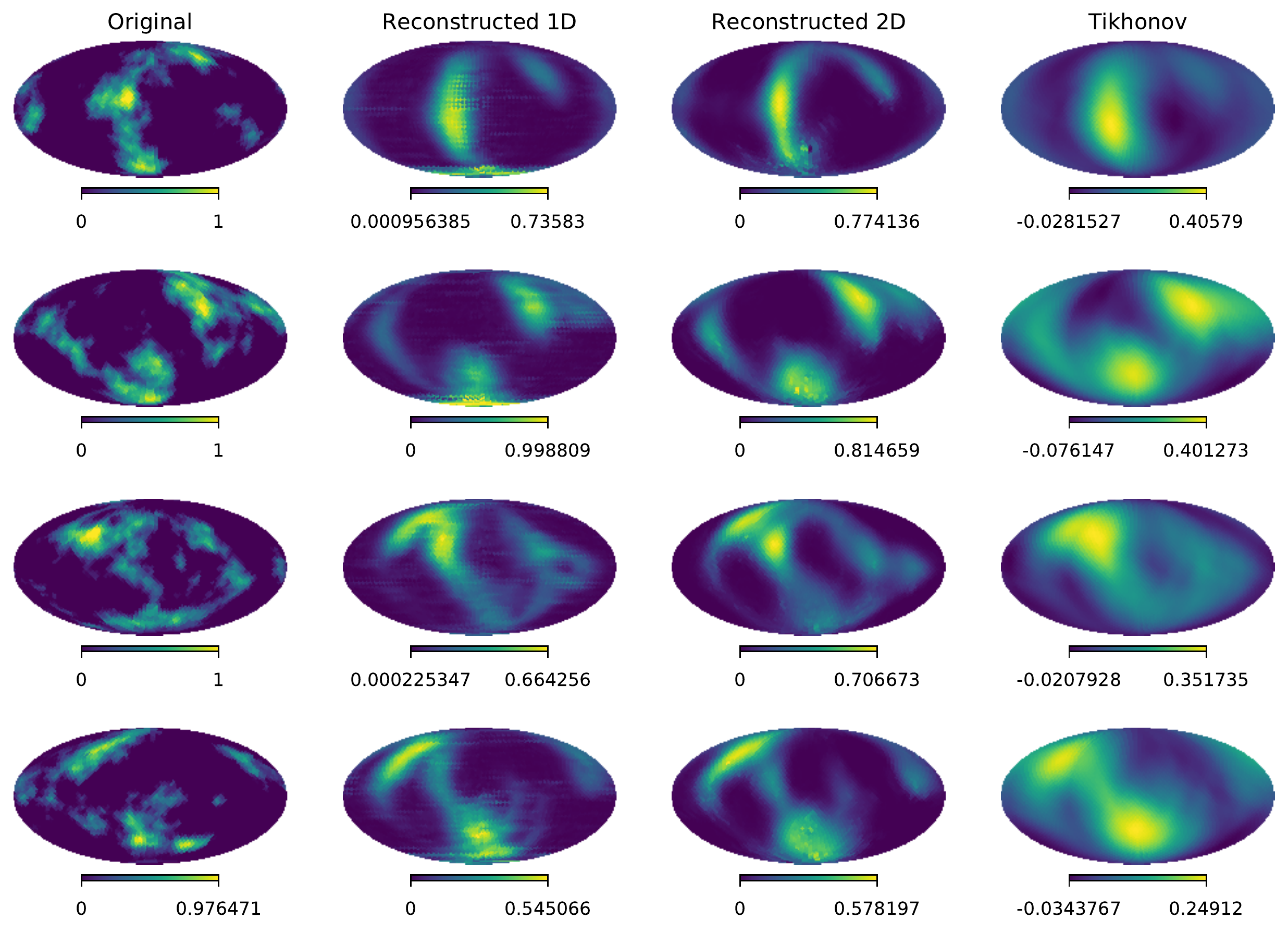}
    \caption{Neural and Tikhonov reconstructions of four different mock cloudless planets observed at the
    same 5 epochs as those of the training set. In this case, all of them share $P_\mathrm{rot}=$24 h, 
    $P_\mathrm{orb}$=365 days,
    $\theta_\mathrm{orb}=0^\circ$ and $\zeta=90^\circ$. We show the reconstructions carried out with
    1D and 2D convolutional networks.}
    \label{fig:reconstruction_mock}
\end{figure*}

A second model, slightly less complex, can also be considered. In this model we assume
that the intrinsic albedo of the clouds is fixed to a value $A_c$. The clouds are then characterized 
by the coverage filling factor, $\mathbf{m}_{c;i}$, inside each discrete pixel on the surface of the planet
at each epoch. Under this approximation, the observed light curve is given by:
\begin{equation}
    \mathbf{d}_i = \boldsymbol{\Phi}_i \left( \mathbf{m}_{s} + (A_c - \mathbf{m}_s) \odot \mathbf{m}_{c;i} \right)  + \vect{\epsilon}.\label{eq:generative_clouds2}
\end{equation}
In this case, the iterative scheme is given by:
\begin{align}
    \mathbf{r}_s^{(k)} &= \mathbf{m}_s^{(k-1)} - 2 \rho \rho^{(k)} \sum_{i=1}^{N_e} \left( \mathbf{t}_{i} - 
    \mathbf{t}_{i} \odot \mathbf{m}_{c;i}^{(k-1)} \right) \label{eq:step1}\\
    \mathbf{m}_{c;i}^{(k)} &= \mathbf{m}_{c;i}^{(k-1)} - 2 \rho \rho^{(k)} \mathbf{t}_{i}
    \odot \left(A_c - \mathbf{m}_s^{(k-1)}  \right)\\
    \mathbf{m}_s^{(k)} &= F^{(k)} (\mathbf{r}_s^{(k)}) \label{eq:step2}.
\end{align}
where
\begin{align}
    \mathbf{t}_{i} &= \boldsymbol{\Phi}_i^T \left[ \boldsymbol{\Phi}_i \left(\mathbf{m}_{s}^{(k-1)} + 
    (A_c - \mathbf{m}_{s}^{(k-1)}) \odot \mathbf{m}_{c;i}^{(k-1)} \right) - \mathbf{d} \right].
\end{align}
All results shown in this paper are obtained with the first model but we describe the second one
here for future reference.

\section{Training}
\label{sec:training}

\subsection{Training dataset}
The neural network is trained with simulated light curves of exoplanets in many different 
configurations. The main ingredients that we need to simulate are the $\boldsymbol{\Phi}$ matrices, which
only depend on the geometry of the problem, and the surface $\mathbf{m}_s$ and cloud albedos $\mathbf{m}_c$, which changes from
planet to planet. The training set is then formed by the triplet $(\mathbf{m}, \mathbf{d}, \boldsymbol{\Phi})$ 
for each example in the training database for cloudless planets. Likewise, for cloudy planets we have 
the quadruple $(\mathbf{m}_s, \mathbf{m}_c, \mathbf{d}, \boldsymbol{\Phi})$ considering all
observed epochs in which the cloud coverage changes.

For the definition of the planetary surface and cloud albedos we use the Hierarchical Equal Area and isoLatitute Pixelization
(HEALPix\footnote{\texttt{http:///healpix.sourceforge.net}}). HEALPix has some desirable 
properties (equal area pixels, efficient available software,\ldots) and has become a standard for the 
pixelization of the sphere, especially in astrophysics \citep[e.g.,][]{2020A&A...641A...1P}. 
For this work we fix $N_\mathrm{side}=16$, which 
results into 3072 pixels on the surface of the planet. 

Geometric kernels are used by our neural approach consistently following the
ISTA iterative scheme so that they do not need to be learned by the neural
network. Therefore, we can get very good generalizations to arbitrary geometries by only 
simulating a sufficient amount of possible geometries and obsering epochs in the training set. More crucial
is the process for the generation of the surface albedos. This is precisely the information
that is exploited by the neural network to \emph{denoise} each step of the iterative scheme and
project the solution on the set of compatible solutions. 

\begin{figure*}
    \centering
    \includegraphics[width=0.9\textwidth]{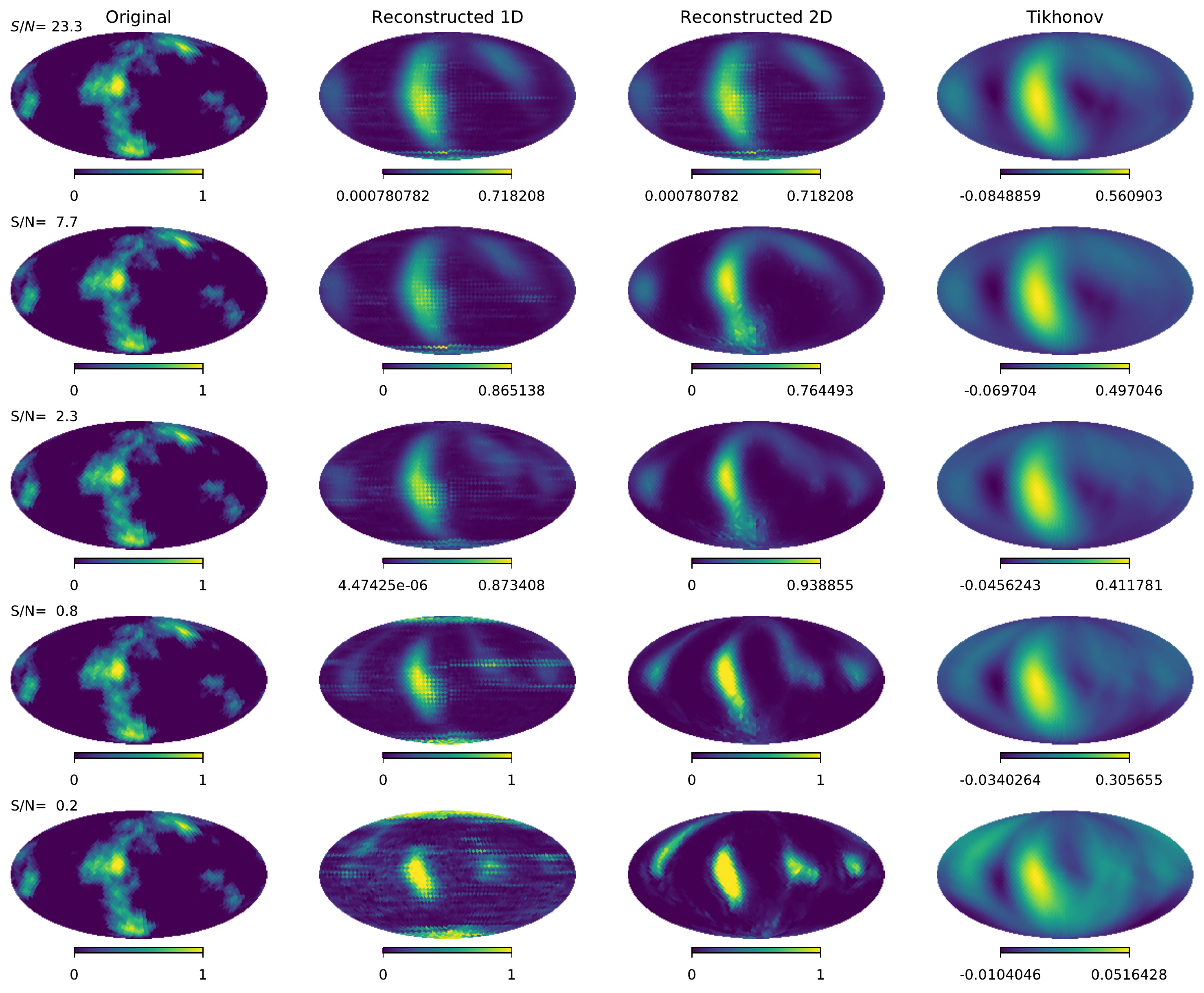}
    \caption{Noise sensitivity of the reconstruction of a mock planet. The geometric parameters are the same
    as those in Fig. \ref{fig:reconstruction_mock} and the signal-to-noise ratio of the light curve is shown in 
    each case.}
    \label{fig:noise_influence}
\end{figure*}

We simulate 5000 different $\boldsymbol{\Phi}$ matrices by randomly setting the rotational
period from 10 to 50 days, the orbital period from 100 to 500 days, and the obliquity and inclination
of the orbit from 0 to $\pi/2$. 
The parameters $\phi_\mathrm{rot}=\phi_\mathrm{orb}=\pi$ are kept fixed given their
arbitrary character. Although one should consider varying them in suitable ranges to cover a good variety of 
potential $\boldsymbol{\Phi}$ matrices, we think that this covers a good amount of potential
planet candidates observations. For a full description of the $\boldsymbol{\Phi}$ matrix requires
the definition of the specific observing times of each epoch.
We assume we observe the planet in 5 epochs that are fixed as 
multiples of the rotation period, in particular $(30,60,150,210,250)$. Each epoch consists of observations in 
steps of 1 hour during an observing period of 24 hours. During this 24 hours interval, global clouds maps are fixed to their daily mean. All $\boldsymbol{\Phi}$ matrices
are computed with the \texttt{exocartographer} software package \citep{2018AJ....156..146F}, using 
the HEALPix pixelization.

The generation of mock exoplanet surfaces is crucial for the success of the algorithm.
The neural network will produce overly simple albedo surfaces if one feeds it with 
simple surfaces. Therefore, we need to 
generate sufficiently complex mock albedo surfaces that can look like potential exo-Earths.
One option is to use the Gaussian process generation from \cite{2018AJ....156..146F}. However,
inspired by the computer graphics procedural generation of planetary surfaces, we prefer
to use 3D Perlin noise \citep{perlin85}. We use the open source
\emph{libnoise} library \footnote{\texttt{http://libnoise.sourceforge.net/index.html}} that
generates correlated noise in the surface of the sphere to produce a database
of mock procedurally-generated planetary surface and cloud albedos. Three examples of
the generated surface and cloud albedos are shown in Fig. \ref{fig:planets}, together
with the final albedo computed with Eq. (\ref{eq:generative_clouds}). We generate 5000 pairs of mock albedo 
surfaces and clouds for the training
by modifying the seed of the Perlin noise generator. The surface albedo contains large zones of 
low albedo which can be compatible with oceans.

\begin{figure*}
    \centering
    \includegraphics[width=0.9\textwidth]{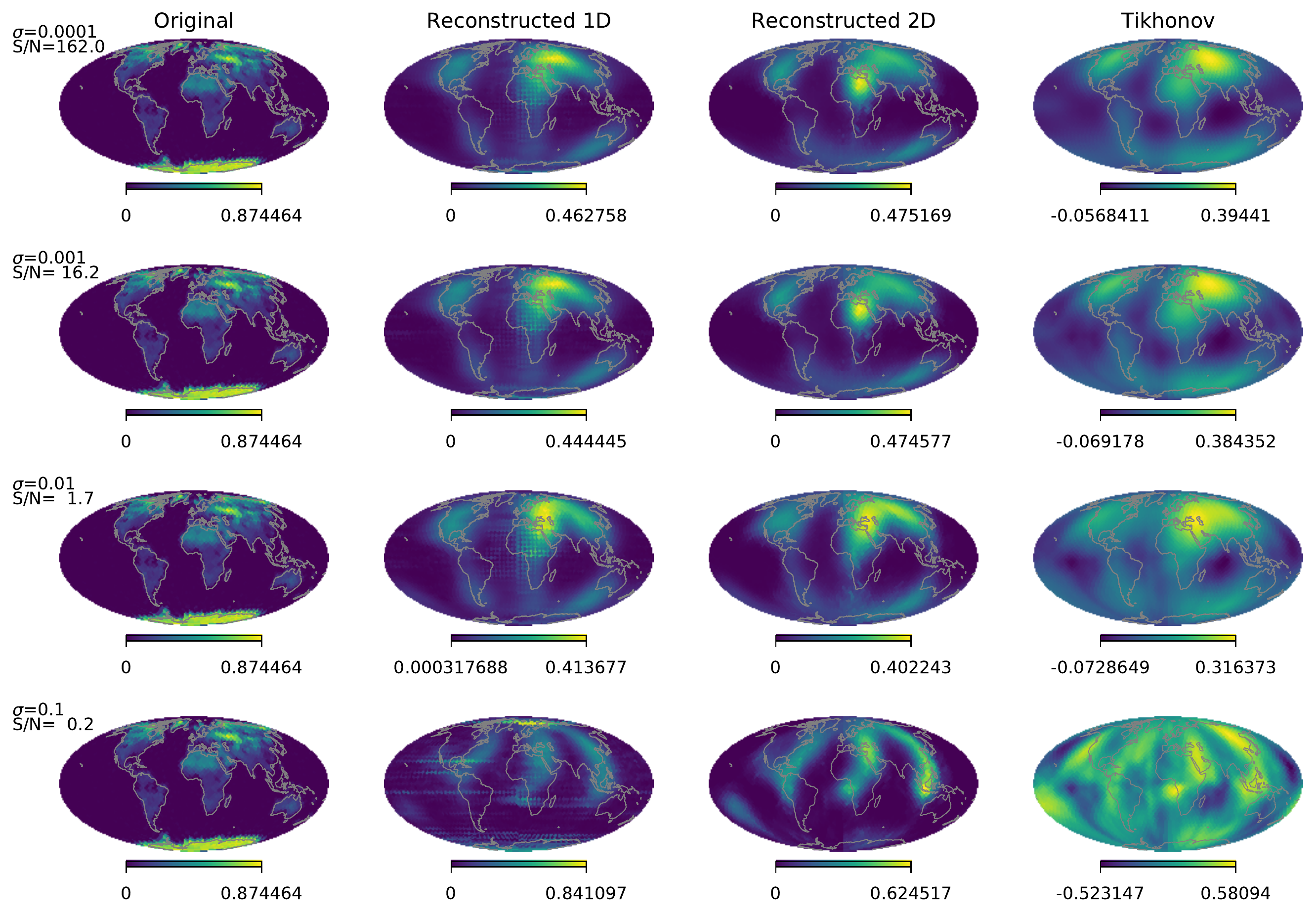}
    \caption{Reconstruction of the cloudless Earth at different $S/N$. The parameter $\sigma$ refers
    to the standard deviation of the noise added to the light curve. The coastlines are shown
    for reference.}
    \label{fig:earth_cloudless}
\end{figure*}

\subsection{Training the neural network}
The training proceeds by chunking the training set in batches of size $N_b=128$ cases for the denoiser
that uses one-dimensional convolutions and $N_b=32$ for the denoiser that uses
two-dimensional convolutions in the sphere (the reduction in the batch size comes
from limitations in the GPU memory). These batches are then fed
to the unrolled neural network in sequence. For each batch, our aim is to force the discrepancy between the reconstructed
albedo map and that of the training set to be as small as possible. The loss function to be minimized is:
\begin{equation}
    \mathcal{L} = \sum_{i=1}^{N_b} \sum_{k=1}^N \omega_k \norm{\mathbf{m}_i^{(k)} - \mathbf{m}_i}_2^2.
\end{equation}
Note that we force all phases of the unrolled neural network to produce intermediate solutions that
get close to the albedo map in the training set (with weight $\omega_k$, that we set to $\omega_k=1$), instead 
of just the final output, $\mathbf{m}^{(N)}$. We found that this
produced a slightly faster behavior during training.
The learnable parameter set includes the weights of all the convolutional layers of $F$ as
well as the step sizes $\rho^{(k)}$. We additionally force $\rho^{(k)}$ to be in the interval $[0.1,3]$.

We define and train the unrolled deep neural network in PyTorch 1.6 \citep{pytorch19}, which
can efficiently compute the gradient of the loss with respect to the parameters
using automatic differentiation. We use the
Adam stochastic gradient optimizer \citep{adam14}, with a learning
rate of 3$\times$10$^{-4}$ during 50 epochs.
We found that the chosen learning rate produces 
suitable results and it was kept fixed for all experiments.  
During training, $N_b$ geometrical kernels and albedo surfaces are randomly picked up 
from the available database to construct each batch. A total of 20k simulated light curves are
used in each epoch of the training.
We utilize an NVIDIA RTX 2080 Ti GPU for the computations and the computing time
per epoch during training is $\sim$60 s in the case of 1D convolutions and $\sim$11 min in the
case of 2D convolutions. A set of unseen validation examples is also reconstructed with the
network at the end of each epoch to check for overfitting.
At the end of training we freeze the neural network parameters that produces the smallest loss function in the validation
set for producing the results shown in the next section.

\section{Results}

\subsection{Cloudless mock planets}
We start by analyzing the results when new mock albedo surfaces are used using the two options
that we consider for the denoising operators, i.e., using 1D or 2D convolutions in the sphere. 
Although having fast reconstructions is not a real issue at the moment we point out 
that the reconstruction of a map with 3072 pixels can be done in $\sim$6 ms when 
using 1D convolutions and in $\sim$11 ms when 2D convolutions are used. These reconstruction
times are obtained by making use of the GPU and do not take into
account any memory transfer time between the main memory and that of the GPU or viceversa, which is
often times a limiting factor. The quoted times include the application of the $N=15$ unrolled 
steps, together with the application of the denoising neural networks in each step.
Although we impose the loss function over all steps of the unrolled iterative scheme, the
results we show are those obtained in the last step. 

\begin{figure*}
    \centering
    \includegraphics[width=0.9\textwidth]{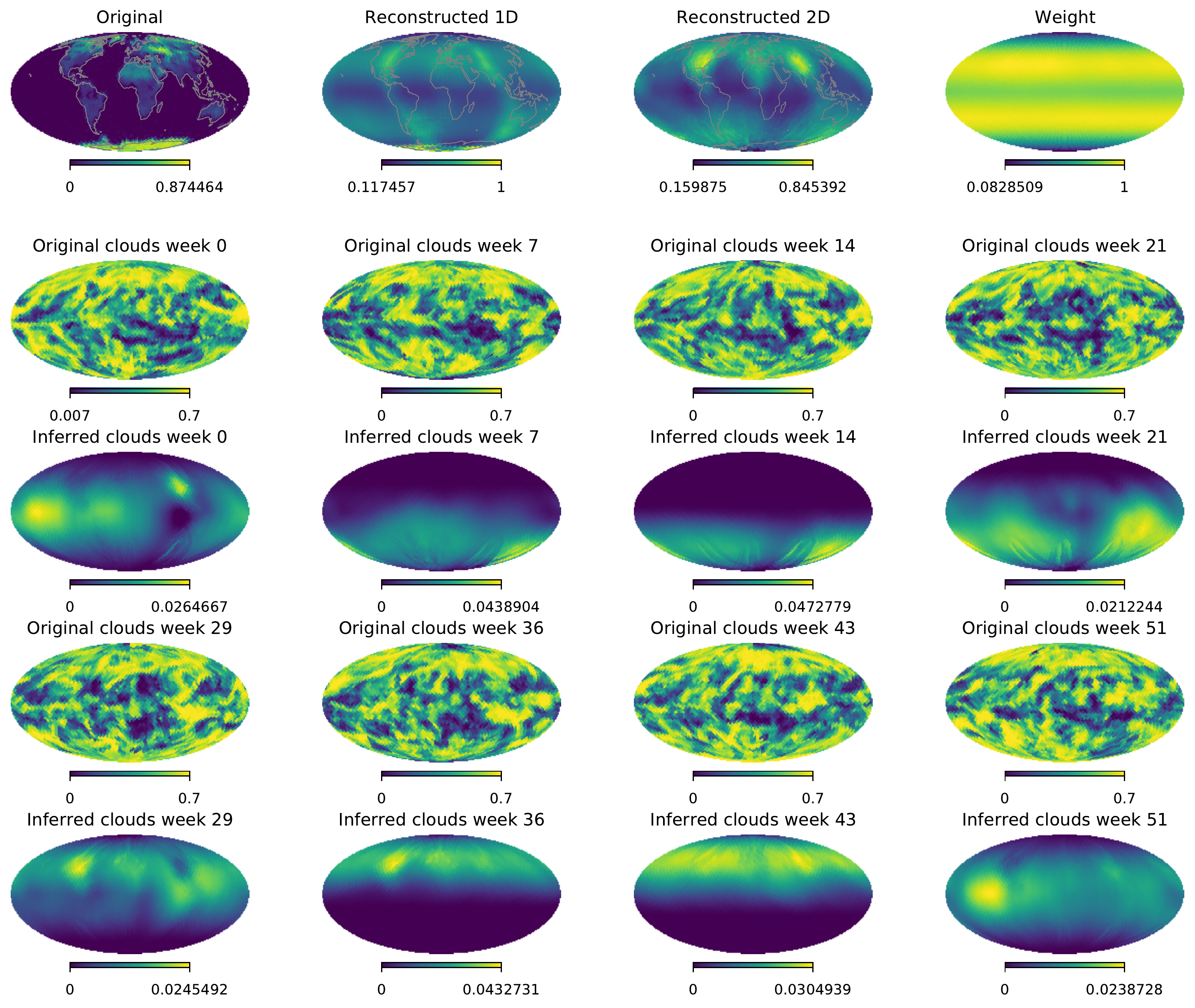}
    \caption{Reconstruction of a cloudy exo-Earth when observed face-on, with 
     $\theta_\mathrm{orb}=0^\circ$ and $\zeta=90^\circ$. The upper row shows the original
     surface albedo map, together with the two reconstructions using different denoisers, as
     well as the relative weight for each element in the surface. The next pairs of
     rows display the original cloud coverage (rows 2 and 4), together with the inferred 
     cloud albedo (rows 3 and 5).}
    \label{fig:earth_faceon}
\end{figure*}

The reconstruction of four mock planets are displayed in Fig. \ref{fig:reconstruction_mock}. The parameters 
are $P_\mathrm{rot}=$24 h, $P_\mathrm{orb}$=365 days,
$\theta_\mathrm{orb}=0^\circ$ and $\zeta=90^\circ$. This corresponds to a face-on observation of
the star-planet system, with the planet spinning axis contained in the ecliptic plane.
This configuration, although probably not dynamically stable, allows us to see the planet
continuously spinning, while the illuminated hemisphere changes during the orbital period. 
We also chose this configuration for easy comparison with previous works that often made 
used of it \citep[e.g.,][]{2011ApJ...739L..62K,2020ApJ...900...48K}.
The planet is observed at the same 5 epochs used during training, which turns out to be 
a very sparse sampling of the light curve. The noise injected 
on the light curve is such that we end up
with a $S/N=2.3$. The left column shows the original map, the 
middle columns shows the two neural reconstructed albedo maps, while the last column displays the
map inferred with the Tikhonov regularization with $\lambda=0.5$ and using the 
optimal value of $\rho$. The value of $\lambda$ was found by trial-and-error looking for the
most contrasted reconstruction without visible artifacts. Both neural reconstructions
show a quite remarkable resemblance with the original map except for some of the high
spatial frequency variations. We note that the artifacts in the 1D regularizer comes from
the fact of not imposing any spatial coherence during the regularization. One of the most
conspicuous artifacts is a checkerboard-like structure that disappears when using the
2D regularizer. All our experiments suggest that the 2D regularizer should be taken
as the baseline.

The neural reconstruction produces much more compact large albedo
regions, in closer agreement with the original maps. However, both reconstructions are unable to reach the
amplitude of the largest albedo regions. Likewise, the very low or zero albedo regions in the original map are reconstructed
with slightly larger albedos, which are used to compensate for the large albedo regions.
It is clear that the neural reconstruction leads to a much better representation of the
albedo map than the Tikhonov-regularized one. Interestingly, we find no visible artifacts 
on the reconstruction in the poles, even though the observational configuration has no access to them.

The neural reconstruction is very resilient to the presence of noise in the light curve,
as shown in Fig. \ref{fig:noise_influence} for a single mock surface. The geometric 
parameters are kept fixed and equal to those
of Fig. \ref{fig:reconstruction_mock}. Increasingly larger amounts of noise are injected
in the observed light curve, giving rise to a broad variation in the $S/N$ ratio, from
0.2 to 23. The value of the regularization parameter in the Tikhonov reconstructions has 
been optimized, by trial-and-error, to produce the best result possible. It is obvious that
the Tikhonov regularization struggles to reconstruct the low albedo regions when 
the noise is very large, while the neural approach consistently recovers a very robust map, even for
very small $S/N$. When the $S/N$ is very low, both the neural and Tikhonov reconstructions
display regions with large albedos that were not present in the original surface.

\begin{figure*}
    \centering
    \includegraphics[width=2\columnwidth]{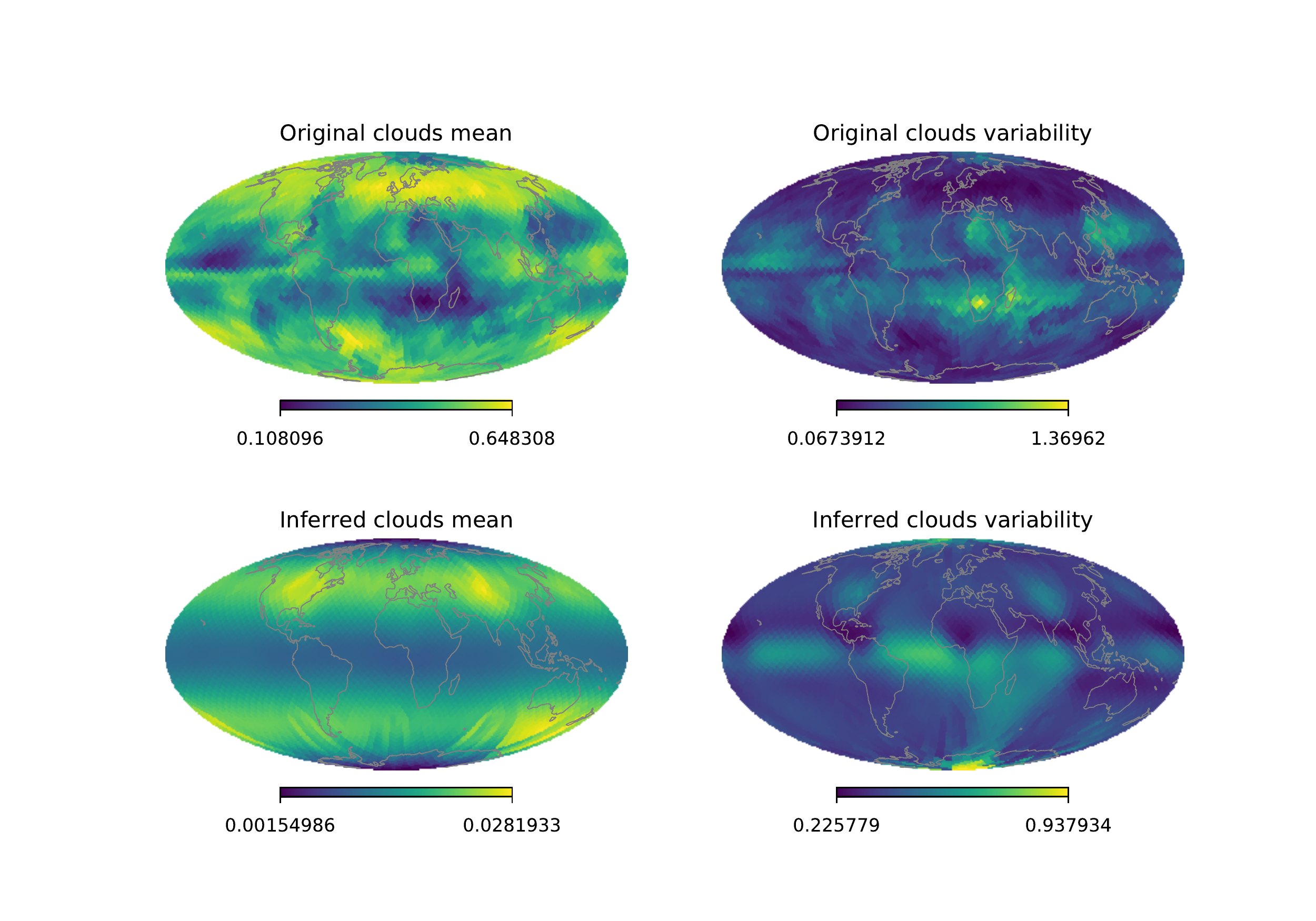}
    \caption{2004 Yearly-mean cloud cover on Earth (top left) and an maps of its variability (top right).  In the bottom panels are the same recovered maps for a simulated Earth, when observed face-on, with $\theta_\mathrm{orb}=0^\circ$ and $\zeta=90^\circ$.}
    \label{fig:statistics_clouds}
\end{figure*}

\subsection{Cloudless Earth}
As an example of the reconstruction of a cloudless planet we show the 
reconstruction of the Earth albedo. The albedo map that we use corresponds to February 2017, obtained 
from the NASA Earth Observation (NEO)
webpage\footnote{See \texttt{https://neo.sci.gsfc.nasa.gov}.}.
We simulate the observation of the Earth for $\theta_\mathrm{orb}=0^\circ$ and $\zeta=90^\circ$.
The light curve is observed for one orbital period (365.25 days) with a cadence of
5 hours, producing a total of $\sim1750$ points in the light curve. We assume that the observations 
are instantaneous, neglecting the smearing effect
produced by the rotation. Both neural reconstructions displayed in Fig. \ref{fig:earth_cloudless}
produce very reliable maps of the surface albedo of the Earth. Almost all continents (North America, South
America, Europe, Asia and Africa) are clearly recovered, with their fine structure in the
albedo. The large albedo of the Sahara desert, central Asia and the northern
part of North America are very well recovered even when the $S/N \sim 1$, with almost the
same extension as in the original surface albedo. Antarctica is not recovered properly 
because of the limited access to the poles in this configuration. Additionally, Australia is 
recovered at a different latitude.

On the contrary, the Tikhonov regularized maps show very
smooth continents. The reconstruction is very robust to the presence of noise
on the light curve. Even when the signal is buried in the noise, with a very low $S/N=0.2$,
the reconstruction is very robust. Some details are lost and in some cases artifacts
appear. However, a comparison with what be achieved with Tikhonov shows the potential
for the learned regularization.

\subsection{Cloudy Earth}
We now turn out attention to the difficult problem of mapping cloudy planets. We analyze
whether the surface of an exo-Earth and the clouds can be mapped with our neural approach.
To this end, we assume that we observe the Earth during a year with a cadence of 5 hours with
a $S/N=16$. We use the
same surface albedo that was used for the cloudless Earth mapping of Fig. \ref{fig:earth_cloudless}. 
The Earth's daily-averaged cloud cover maps are obtained from the daily 20-year database of the International Satellite Cloud Climatology Project\footnote{See \texttt{https://isccp.giss.nasa.gov/}} \citep{Rossow2019}. We arbitrarily chose to use the data from year 2004.
Even though the database contains a different cloud cover map each day, we make the simplifying
assumption that the cloud covering is stable for one week. Since the typical lifetime of large-scale cloud pattern on earth is about 1-week \citep{2008ApJ...676.1319P}, this makes the problem slightly
better constrained. 

We consider two observational geometries, both of them allowing a potentially complete mapping
of the surface. The first one is a face-on observation ($\theta_\mathrm{orb}=0^\circ$) with 
obliquity $\zeta=90^\circ$. This configuration, similar to that used in the previous 
examples, produces an illuminated hemisphere that changes smoothly with the orbit from
the north pole to the south pole and viceversa, while the spinning modulation gives information 
about the structures in the illuminated hemisphere. The second case is an
edge-on configuration ($\theta_\mathrm{orb}=90^\circ$) with obliquity $\zeta=23^\circ$
without considering the eclipse. This configuration is less informative because the 
observations are much more sensitive to the equatorial and tropical regions than to
those of larger latitude, but it is a much more likely scenarios for real exoplanet observations.

\begin{figure*}
    \centering
    \includegraphics[width=0.9\textwidth]{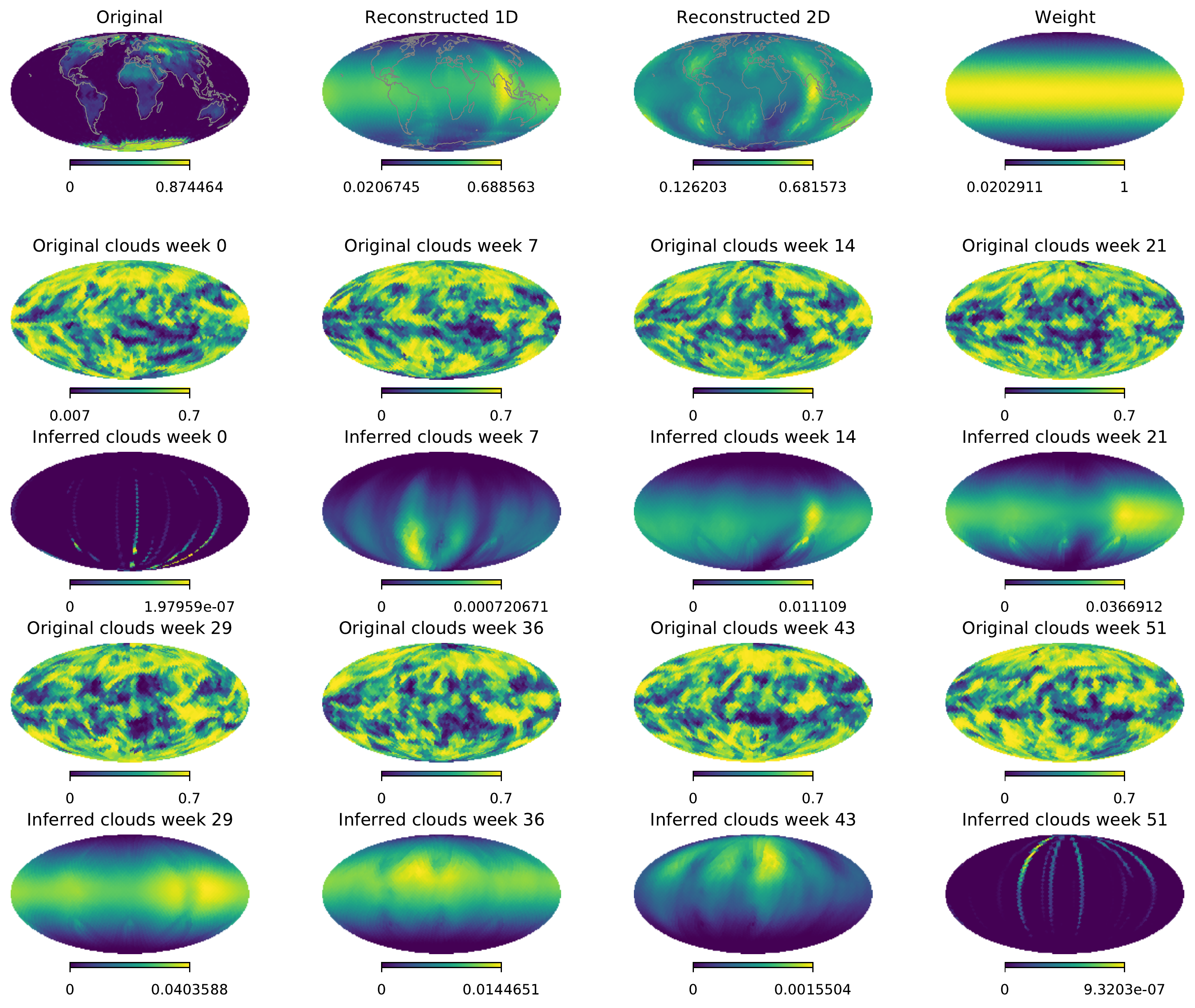}
    \caption{Like Fig. \ref{fig:earth_faceon} but for edge-on observation with 
    $\theta_\mathrm{orb}=90^\circ$ and $\zeta=23^\circ$.}
    \label{fig:earth_edgeon}
\end{figure*}

Figure \ref{fig:earth_faceon} shows the results of applying the neural reconstruction
using the two denoisers. The upper row displays the original map, the two reconstructed
maps, as well as the weight that this configuration puts on every point on the surface.
The remaining rows display the original cloud albedo (rows 1 and 3) and the inferred 
cloud albedo (rows 2 and 4) for 8 epochs along the year of observation (labeled with the
specific week). When compared with Fig. \ref{fig:earth_cloudless}, the inferred surface 
albedos are strongly affected by the presence of the clouds. The large albedo of
North America is correctly recovered, but the Sahara desert and
the Central Asia high albedo regions are slightly displaced in latitude, probably
as a consequence of the persistent clouds, as explained in the following. We have checked
that this displacement is reduced and the maps converge towards those of Fig. \ref{fig:earth_cloudless} 
when the amount of clouds is decreased. 

The cloud albedo on each epoch clearly shows 
the partial information produced by the change in the illuminated hemisphere. We note
that the amplitude of the inferred albedo of the clouds is systematically very low.
This is probably a consequence of the weak dependence of the total albedo on the
specific albedo of the clouds when the surface albedo is large. For this reason, although
the learned scheme tends to produce small albedos for the clouds, one can scale them
up while still fitting the observations. As a consequence, obtaining absolute cloud
albedos seems to be not possible with monochromatic observations.

Even if we cannot obtain the cloud map per epoch, it is worth analyzing if one can 
recover the mean cloud coverage and, more interestingly, a measure of its variability.
The upper row of Fig. \ref{fig:statistics_clouds} shows the 2004 yearly-mean cloud coverage 
for Earth and its relative variability (defined as the standard deviation normalized by the mean). The most
conspicuous features are persistent clouds close to the east coast of South America, in the northern
part of America, Europe and Asia and in the southern coast of Australia. These are in 
good agreement with some of the regions of large albedo obtained in the neural reconstructions.
Therefore, we conclude that these reconstructions are able to obtain a combination of
the surface albedo, together with the persistent clouds. Additionally, the recovered 
cloud relative variability (lower row) nicely correlates with the one measured on Earth. Large 
cloud relative variabilities are found close to the Equator, in southern Africa and
in part of eastern Asia. These are areas with strong daily and seasonal cloud variability. Consequently, 
we conclude that this mapping method is able
to obtain a roughly estimate of cloud variability on an exo-Earth. When applied to rocky planets in the 
habitable zone of their host stars, the bulk atmospheric composition from spectroscopic studies as well as 
the measured equilibrium temperature of the planet will be able to constrain and associate this 
cloud variability to that of the water cycle, implying not only water vapor in the planetary 
atmosphere but also significant amounts of liquid water in the surface.

Finally, we show in Fig. \ref{fig:earth_edgeon} the reconstructions in the edge-on observation. 
Unfortunately, the limited latitude sensitivity of this configuration introduces an ambiguity in 
latitude of all structures that cannot be overcome. Still the recovered ``D-shaped'' maps capture well 
the the three major continental features of the original maps (Americas, Europe/Africa and Asia) and 
thus the inhomogeneity of the planetary surface. We have been unable to significantly improve the
quality of the mapping by increasing the duration of the light curve. As a consequence of the poor
mapping of the surface, the cloud mapping is also defective and the cloud variability is also
poorly inferred.

\section{Conclusions}
We have presented a mapping technique for exo-Earths using neural networks as 
regularizers\footnote{\texttt{https://github.com/aasensio/neural\_exocartography}}. The standard iterative
scheme for the solution of linear problems with regularization is unrolled as a neural network and trained
in an end-to-end fashion. The best regularizer is found to be a convolutional neural network acting
on the surface of the planet. Given the absence of observed maps of exo-Earths, we use albedo surfaces
generated procedurally trying to mimick our planet. The mapping technique is general and 
can be re-trained in case better options for albedo surfaces are found in the future.

We apply the method to cloudless planets, showing that the surface of such planets can be 
very well recovered with our methodology. The position of the continents and their extension shows
a clear improvement with respect to more standard regularizations like Tikhonov. Additionally, the results 
are extremely resilient to the presence of noise on the light curve. We also show how cloudy
planets can also be roughly mapped with the same method using only monochromatic observations. The inferred
albedo maps contain information about a combination of the surface albedo and persistent clouds, which cannot
be disentangled. This degeneracy can potentially be broken if polychromatic observations are used if 
clouds and surface albedos have a different behavior with wavelength \citep{Cowan2009}.

Finally, although detailed cloud mapping for each observing epoch cannot be accurately retrieved, we show that we can obtain a good estimation of the mean persistent cloud distribution and, more importantly, their map of variability.
If applied to rocky planets in the habitable zone of their stars, this cloud variability maps, 
together with the measured bulk atmospheric composition and equilibrium temperature of the 
planet, will be able to constrain the presence of a water cycle in the planetary atmosphere/surface. 
Thus, our method implies the potential to empirically infer the presence of significant amounts of 
liquid water in the planet's surface.

\begin{acknowledgements}
We acknowledge financial support from the Spanish Ministerio de Ciencia, Innovaci\'on y Universidades through project PGC2018-102108-B-I00 and FEDER funds. 
This research has made use of NASA's Astrophysics Data System Bibliographic Services.
We acknowledge the community effort devoted to the development of the following 
open-source packages that were
used in this work: \texttt{numpy} \citep[\texttt{numpy.org},][]{numpy20}, 
\texttt{matplotlib} \citep[\texttt{matplotlib.org},][]{matplotlib}, \texttt{PyTorch} 
\citep[\texttt{pytorch.org},][]{pytorch19} and \texttt{zarr} (\texttt{github.com/zarr-developers/zarr-python}).
\end{acknowledgements}

%
%



\end{document}